\documentclass[letterpaper, 10 pt, conference]{ieeeconf}  
\IEEEoverridecommandlockouts                              
\usepackage{graphics} 
\usepackage{epsfig} 
\usepackage{mathptmx} 
\usepackage{times} 
\usepackage{amsmath} 
\usepackage{amssymb}  
\usepackage{color}
\usepackage{booktabs}
\usepackage{verbatim}
\usepackage{float}

\title
{\LARGE \bf Drone-Assisted Communications for Remote Areas and Disaster Relief}
\author{Anousheh Gholami, Usman A. Fiaz, and John S. Baras\\
\tt\small Department of Electrical and Computer Engineering\\
Institute for Systems Research\\
University of Maryland, College Park, MD 20742, USA\\
\{ \tt\small anousheh | fiaz | baras @umd.edu \}
}

\begin{document}

\maketitle
\thispagestyle{empty}
\pagestyle{empty}
\section{Background and Motivation}
Recent years have seen a great deal of interest in deploying unmanned aerial vehicles (UAVs), also known as drones, for applications such as aerial transport of goods, search and rescue, precision agriculture, wildlife conservation, real-time monitoring, security and surveillance. Besides this wide range of conventional applications of drones, UAV-assisted communication systems have emerged as one of the key technologies for the development and expansion of fifth generation (5G) networks, and have therefore attracted a lot of interest from both academia and industry \cite{naqvi2018drone}.

UAV-assisted wireless communications can be categorized into three typical use cases (as in \cite{zeng2016wireless}):
 \begin{itemize}
\item \textit{UAV-assisted ubiquitous coverage} where the UAVs are deployed to support existing communication infrastructure to provide seamless wireless coverage; for example, in crowded concerts and in sport events at massive stadiums.
\item \textit{UAV-assisted relaying} for connecting distant users or user groups without direct and/or reliable communication link. In the case of disaster-relief operations, deployment of UAVs as relay nodes could be used to establish communication between disconnected clusters of ground users (for example, the front-line disaster response teams and the base-station). 
\item \textit{UAV-assisted information dissemination and data collection} where the UAVs are deployed to collect/disseminate information to and/or from a large set of distributed wireless devices, for example, supporting the Internet of Things (IoT) communications, or wireless sensor networks. 
 \end{itemize} 
 

In comparison with the existing terrestrial infrastructure, UAV-assisted communication systems are faster to deploy and are more cost-effective. In addition, these systems are highly adapting towards new environments due to controllable mobility of UAVs in 3D space and are also likely to yield better communication channels because of their short-range Line-of-Sight (LoS) links.
However, deployment of UAVs as flying base stations is much more challenging than the existing ground cellular systems. This difficulty arises from the limited flight time of UAVs, constraints on their power consumption and high maneuverability, and the problem of providing reliable backhaul connections.

There are several existing works in literature which focus on an optimal UAV deployment scheme considering only the air-to-ground (A2G) communication links between the drones and the ground users (GUs). For instance, in \cite{bor2016efficient, al2014optimal} and \cite{wu2018joint}, the authors consider different optimization objectives for this problem such as minimum power consumption, maximum user coverage, and maximum minimum-throughput. On the other hand, the problem of multi-hop backhaul connectivity, where air-to-air (A2A) communication links are present, is still by and large an unexplored area of research, and hence it is the focus of our attention. 

\begin{figure}
\begin{center}
\includegraphics[width=8.6cm]{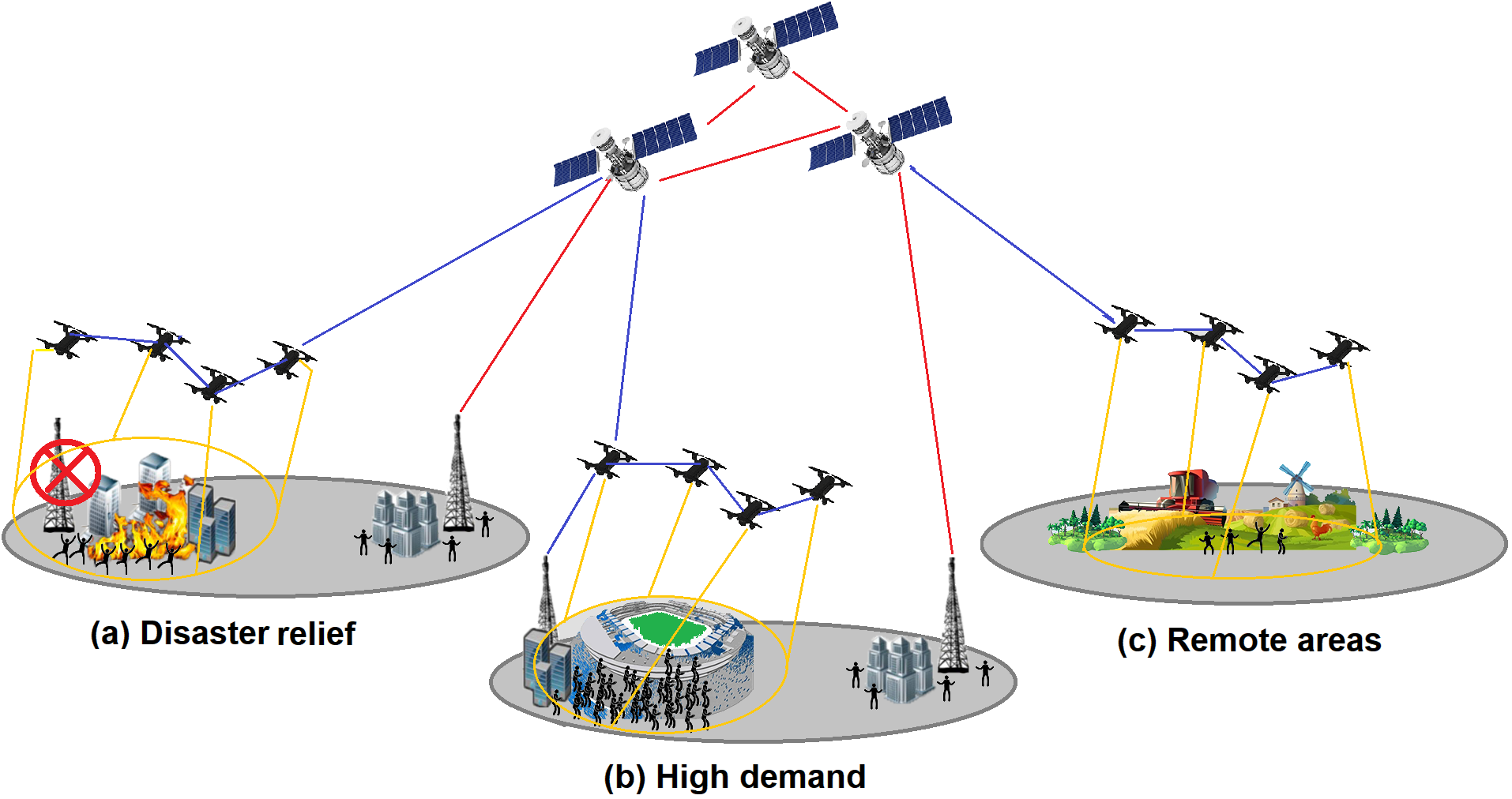}    
\caption{\hspace{-0.35cm}Search and rescue in disaster struck areas, coverage in rural, underdeveloped, and remote areas, and coping with reliable connectivity and high demand in crowded localities are three of the most realistic and important use cases for the deployment of UAV-assisted communication networks.}\vspace{-0.5cm}
\label{fig:adhoc} 
\end{center}
\end{figure}

\section{Objectives}

Our intention is to design and analyze the access and backhaul connectivity of a drone-assisted communication network, while minimizing the resources required i.e., the number of UAVs. 

A work along similar lines is presented in \cite{li2018placement}, where the authors maximize the minimum throughput among all the GUs subject to the flow conservation constraints on the UAVs. The optimization metric is the UAVs’ deployment locations, along with the bandwidth and power allocation of the access (A2G) and backhaul (A2A) links. However, this work only considers the downlink communications. 

We explore an end-to-end (including access and backhaul links) UAV-assisted wireless communication system, considering both uplink and downlink traffics, with the goal of supporting demand of the GUs using the minimum number of UAVs. Moreover, in order to extend the operational (flight) time of UAVs, we exploit an energy-aware routing scheme.


\section{The Proposed Approach}

We consider a ground network of users with a topology represented by a graph $G = (\mathcal{N}, \mathcal{L})$. The set of links $\mathcal{L}$ between GUs is assumed given, based on a neighborhood discovery mechanism and the matrix $D$ is the traffic demand of origin-destination (OD) pairs. The objective is to deploy a network of UAVs to support the traffic demand of the ground network in an energy efficient manner. We decompose this design problem into three phases. 

In the first phase, the GUs are grouped using a weighted k-means clustering approach (with density-based-box initialization) where the value of k is determined based on $D$ and GUs locations $w_n=(x_n,y_n) ,n=1,..,N$. The weight of each node is its total traffic demand. 
The goal of the second phase is to find the minimum number of UAVs $L$ and their projected locations on the ground $a_l = (x_l^a,y_l^a), l = 1,...,L $, such that the resulting network is connected and the demand of GUs is satisfied. We assume that all UAVs have a fixed flying altitude $h$. The UAVs are introduced in two steps. First, each weighted centroid of the ground clusters is assigned to an UAV location. We use a merging algorithm to move the UAV locations until the UAVs with a total demand less than $C_{max}$ are merged, while staying connected to their associated GU clusters. Secondly, another set of UAVs is introduced in the case that existing network is not connected. This step is done using Deterministic Annealing (DA) \cite{rose1998deterministic} similar to the approach proposed in \cite{perumal2008aerial}.
In the third and final phase, we aim to extend the operation of UAVs through an energy-aware routing. This problem is formulated as a Mixed Integer Linear Program (MILP), with the objective of maximizing supported traffic and minimizing total power consumption.

\section{Results and Prospects}

\begin{figure}
\begin{center}
\includegraphics[width=8.0cm]{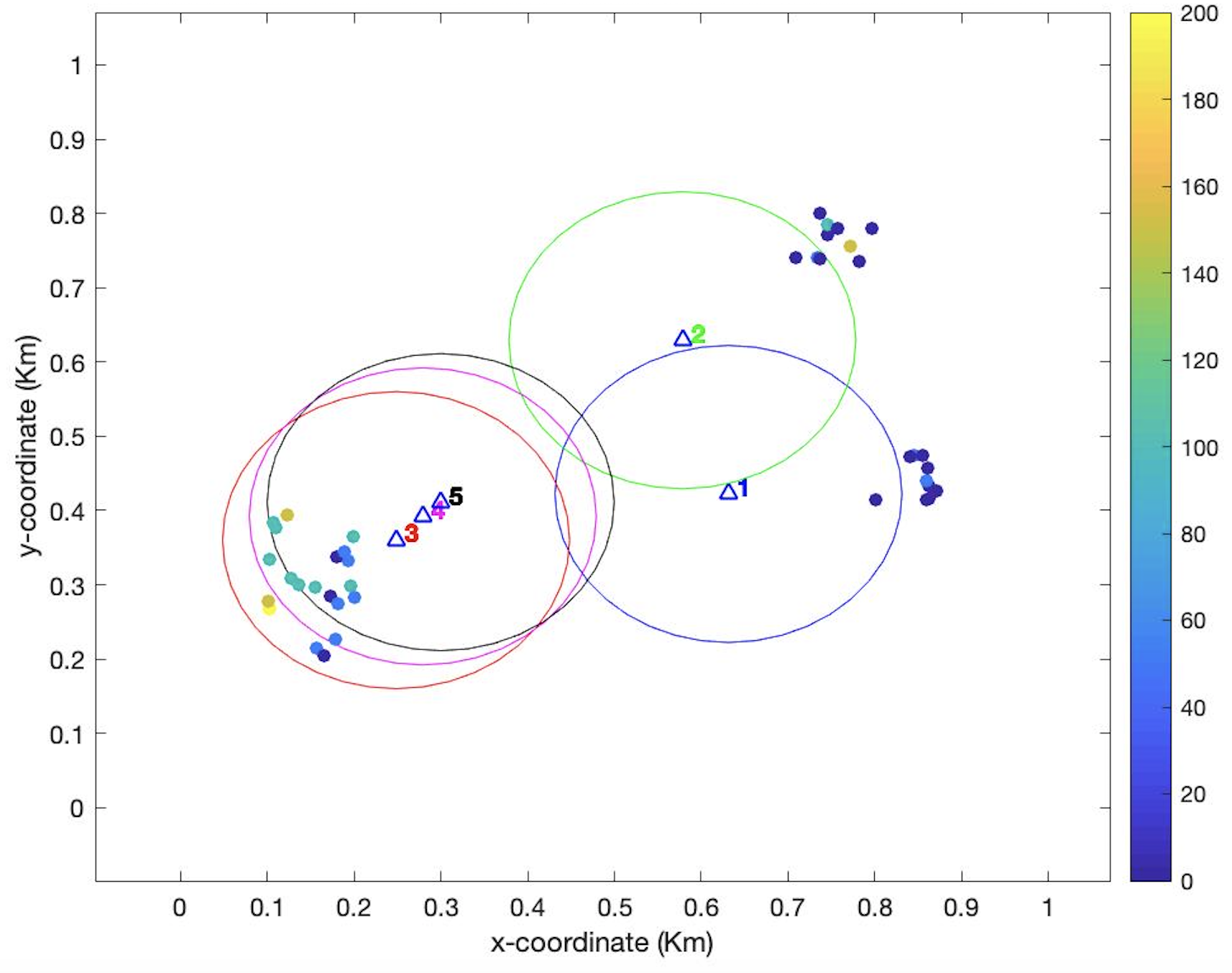}    
\caption{An example of UAV placement}\vspace{-0.4cm}
\label{fig:UAV placememtn} 
\end{center}
\end{figure}

\begin{figure}
\begin{center}
\includegraphics[width=8.0cm]{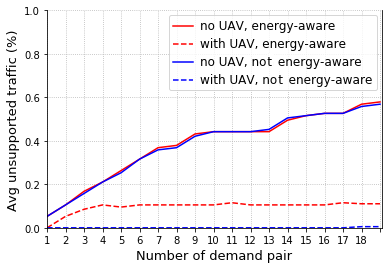}    
\caption{Comparison of the unsupported traffic demand}\vspace{-0.25cm}
\label{fig:eta} 
\end{center}
\end{figure}

In this section, we demonstrate some results of the proposed approach, and the significance of using UAVs to provide an enhanced coverage in areas with limited or no infrastructure.
The ground network consists of 40 nodes distributed in three spatial clusters where the nodes in each cluster construct a connected graph and the nodes in two different clusters cannot communicate directly (see Fig.~\ref{fig:UAV placememtn}). An instance of the proposed UAV placement algorithm is depicted in Figure \ref{fig:UAV placememtn}. The GUs are showed by points in the 2D plane with the colors standing for the user traffic demand (in Kbps). The UAVs are placed at the center of the circles which show the UAVs communication range. 
Figure \ref{fig:eta} illustrates the average unsupported traffic defined as the fraction of network total traffic demand which is not fulfilled. It shows that the average unsupported traffic demand increases as the number of OD pairs increases in the network without UAVs. Whereas, the deployment of flying bases stations keeps this fraction close to zero. Moreover, the comparison of power consumption of UAVs in the case of energy-aware routing with the one with no energy-aware routing reveals a trade-off between energy efficiency and fraction of supported traffic, which is expected. 

In future work, the time-varying deployment of UAVs is considered when the ground network topology is changing and UAVs maximum speed is constrained.

\bibliography{IEEEabrv,ieeeconf}


\end{document}